\documentstyle[preprint,prl,aps]{revtex}
\begin{document}
\draft
\preprint{}

\title
{Transport and magnetic properties of La$_{1-x}$Ca$_x$VO$_3$}
\author{K. Maiti, N. Y. Vasanthacharya and D. D. Sarma$^*$} 
\address
{Solid State and Structural Chemistry Unit,
Indian Institute of Science,
Bangalore 560 012,
INDIA}
\date{\today}
\maketitle

\begin{abstract}
We report the temperature dependence of transport and magnetic
properties of La$_{1-x}$Ca$_x$VO$_3$ for $x$ = 0.0, 0.1, 0.2, 0.3,
0.4 and 0.5.  The system exhibits an insulator-to-metal transition
concomitant with an antiferromagnetic-to-paramagnetic transition near
$x$ = 0.2 with increasing substitution. Disorder effects are found to
influence the low temperature transport properties of both insulating
and metallic compositions near the critical concentration. At higher
temperatures, the resistivity of the metallic samples is found to exhibit
either a $T^{1.5}$ or a $T^2$ dependence depending on the
composition.  The molar susceptibility for the metallic samples
indicate substantial enhancements due to electron correlation.
\end{abstract}

\pacs{PACS Numbers: 71.30.+h, 71.27.+a, 75.20.Hr, 75.30.Cr}

\section{Introduction}

The study of metal-insulator transition in early transition metal (Ti
and V) oxides has attracted a lot of attention in recent times
\cite{lasrtio3,str4,mott1,tokura,lacavo,ycavo3,casrvo3,v2o3}. 
Most of these oxides have the perovskite structure which allows for
doping of charge carriers by chemical heterovalent substitutions over
a wide range of compositions without breaking the structural network.
Moreover, this basic structure type makes it possible to tune various
structural parameters, such as the bond angles and bond lengths,
which directly affect the bandwidth by altering the intra- and
inter-cluster hopping interactions. Thus, it becomes possible to
investigate metal-insulator transitions as a function of charge
carrier tuning as well as bandwidth (or correlation effects) control.

Two such series of compounds that are related to the present work are
La$_{1-x}$Sr$_x$VO$_3$ and Y$_{1-x}$Ca$_x$VO$_3$, which have been
recently studied for several values of $x$ \cite{str4,mott1,ycavo3}.
LaVO$_3$ is an antiferromagnetic insulator with N\'{e}el temperature
around 140 K \cite{lvomag}, with a distorted perovskite structure
\cite{str1,str2,str3,str5,str6,str7}. Optical
conductivity measurement \cite{tokura} suggests the charge excitation
gap (the Mott-Hubbard gap) to be about 1.1 eV.  Substitution of
trivalent La with divalent Sr effectively leads to the doping of
holes into the system, thereby driving it to a metallic phase at
about 22\% hole doping \cite{mott}, a value very near to the
percolation threshold ( $x$ $\sim$ 0.26). The system remains
antiferromagnetic throughout the insulating regime ($x$ $\leq$ 0.2)
and shows paramagnetic susceptibility in the metallic phase
\cite{mag}.  The insulator to metal transition in this system
has been believed to be driven by Anderson localization
\cite{mott1,mott2}. Sr substitution in this system drives the system
towards the ideal cubic perovskite structure which is the structure
of the end member SrVO$_3$.  YVO$_3$ is also an antiferromagnetic
insulator with orthorhombically distorted perovskite structure
\cite{stryvo}. Here the V-O-V bond angle is lower than that of
LaVO$_3$ leading to a decreased bandwidth in YVO$_3$. With the
substitution of Y$^{3+}$ by Ca$^{2+}$ it transforms to a metallic
phase, but the minimum amount of hole doping needed to get the
metallic phase is very large ($x$ $\sim$ 0.5) compared to the case of
La$_{1-x}$Sr$_x$VO$_3$ ($x$ $\sim$ 0.22). This large value of $x$ in
YVO$_3$ rules out the possibility of percolation to be one
explanation for the MI transition in this system.  One possible
reason for this large value for the critical concentration might be
the smaller band width in YVO$_3$ compared to LaVO$_3$ discussed
above. This suggests that changing electron correlation effects also
play an important role in these doping dependent MI transitions.

We study the transport and magnetic properties of LaVO$_3$ and its Ca
substituted compositions, La$_{1-x}$Ca$_x$VO$_3$ for various values
of $x$. This system is different from La$_{1-x}$Sr$_x$VO$_3$ since
the effective ionic radius Ca$^{2+}$ (1.34 \AA) is very close to that
of La$^{3+}$ (1.36 \AA), in contrast to that of Sr$^{2+}$ (1.44 \AA)
\cite{shanon}. The crystal structures of the two end members LaVO$_3$
and CaVO$_3$ are both distorted perovskite type
\cite{str4,str1,str2,str3,str5,str6,str7,cvostr}, whereas SrVO$_3$ is cubic 
\cite{svostr}.  Thus, unlike Sr$^{2+}$
substitution, Ca$^{2+}$ substitution in LaVO$_3$ dopes hole states
without much change in structural parameters. In view of this, it is
reasonable to expect that the changes in the transport and magnetic
properties in La$_{1-x}$Ca$_x$VO$_3$ are more controlled by the
tuning of the doping level with comparatively less influence from the
changes in the correlation effects, in contrast to the previously
studied La$_{1-x}$Sr$_x$VO$_3$ and Y$_{1-x}$Ca$_x$VO$_3$.

\section{Experimental}

The samples of La$_{1-x}$Ca$_x$VO$_3$ ($x$ = 0.0, 0.1, 0.2, 0.3, 0.4 and
0.5) have been prepared by taking stoichiometric amounts of predried
La$_2$O$_3$, CaCO$_3$ and V$_2$O$_5$ in appropriate proportions. The
components have been ground together and heated at 600$^o$ C and then
800$^o$ C for 24 hours each. Then the obtained mixture have been
reground and reduced at 800$^o$ C for 24 hours in hydrogen
atmosphere. Finally, the black color powder obtained has been ground
and pelletized, and melted in a dc arc furnace in inert gas
atmosphere. All the samples have been characterized by x-ray powder
diffraction (XRD) pattern obtained from a JEOL-8P x-ray
diffractometer using Cu K$\alpha$ radiation. XRD patterns show single
phases for all the compositions. The structure remains the same, in
the whole composition range prepared.

It is well known that there is always some amount of oxygen
nonstoichiometry in these oxides \cite{tio,lamno,lafeo}. We have
estimated the exact oxygen content in each of the samples by heating
the samples at 500$^o$ C for about 72 hours in flowing oxygen,
thereby converting all V ions to the highest oxidation state of
V$^{5+}$. By monitoring the weight gain in this process, we estimate
that La$_{1-x}$Ca$_x$VO$_{3+\delta}$ samples formed with $\delta$ =
0.04, 0.03, 0.04, 0.01, -0.01 and 0.0 for $x$ = 0.0, 0.1, 0.2, 0.3,
0.4 and 0.5 respectively. Thus it appears that there are some doped
hole states arising from slightly nonstoichiometric compositions in
addition to those introduced by heterovalent substitution of
La$^{3+}$ by Ca$^{2+}$ in these samples, particularly for the samples
with small values of $x$.  As will be shown later in the text, such
oxygen nonstoichiometry influences the low temperature transport
properties of these samples especially within the insulating regime.

The resistivity measurements have been carried out using a dc four
probe resistivity set up following Van der Pauw's method.  The dc
magnetic susceptibility measurements have been performed in the
temperature range 20-300 K using a Lewis coil force magnetometer
at a field of 5 kOe.

\section{Results and Discussion}

We show the resistivity ($\rho$) of LaVO$_3$ as a function of
temperature ($T$) in Fig. 1. The result clearly suggests an
insulating behaviour for this compound. However the transport does
not follow a single activated dependence as illustrated in the inset
where we show the dependence of resistivity ($\rho$) on the
temperature ($T$) in an $ln~\rho ~ vs~ 1/T$ plot. The inset
suggests a simple activated behavior at the high temperature regime,
while there is a considerable deviation for $T < 120 $ K (i.e. $1000/T >$ 8.5).
 In this low
temperature regime, the resistivity appears to be best described by
the variable range hopping mechanism \cite{mott}, as illustrated in
the inset by the linear dependence of $ln~ \rho$ on $1/T^{1/4}$.
Combining these two behaviors, we find that the experimental 
transport data is well described
over the entire temperature range in terms of simultaneous
contributions to the conductivity ($\sigma$) both by an activated
process and by variable range hopping (VRH) in the form :

$$\sigma = \sigma_{01}exp[-{E_g \over {2K_BT}}] +
\sigma_{02}exp[-{T_0 \over {T^{1/4}}}]$$

The resulting best fit is shown by a solid line through the 
data points in the main figure
over the entire temperature range. From the fitting we obtain the
value of the gap ($E_{g}$) to be about 0.12 eV.  It is to be noted here
that this estimate of the transport gap is about one order of
magnitude smaller than the intrinsic Mott-Hubbard gap (1.1 eV) in
LaVO$_3$ estimated from optical spectroscopy
\cite{tokura}. Thus we believe that the smaller gap arises from
doping of hole states above the top of the valence band due to the
inevitable presence of oxygen nonstoichiometry even in the undoped 
sample as already
pointed out in the experimental section. This smaller gap is dominant
for the transport properties at room temperature and below, since
charge excitations across the large intrinsic Mott-Hubbard gap are
suppressed at these temperatures. The small contribution from VRH to
the conduction process at the lowest temperature suggests the 
presence of low but finite
density of states (DOS) localized due to disorder effects
\cite{mott1} at the Fermi energy. 
We discuss further these aspects later in the text.

The resistance of the smallest substituted composition
La$_{0.9}$Ca$_{0.1}$VO$_3$ is plotted as a function of temperature in
Fig. 2. The diverging resistivity with decreasing temperature suggests
a non-metallic ground state of this sample in spite of 0.1 hole doping
in every formula unit. In order to understand the nature of the
non-metallic state in this case, we plot $ln~ \rho$ as a function of
$1/T$ in the inset. The high temperature (approximately $ T > 110$ K 
or $ 1000/T < $ 9) end of this
plot shows an activated behavior as suggested by the straight
line through the data in the inset; however, the resistivity plot
deviates markedly at lower temperatures. In the inset of Fig. 2, it
is seen that the logarithm of resistivity is linear in
$1/T^{1/4}$ over a wide range of temperatures.  These observations
suggest that the transport properties of La$_{0.9}$Ca$_{0.1}$VO$_3$
is dominated by variable range hopping \cite{mott} for $T$ $<$ 110 K,
while an activated process contributes significantly for $T$ $>$
110 K.  In this case also the experimental data can be accurately
described under the assumption that the total conduction process is
contributed simultaneously by variable range hopping and an activated
process over the entire temperature range; this is illustrated by the
solid line through the data in the main frame in Fig. 2,
representing the best fit. This fit yields an estimate of the gap in
the activated charge transport process to be 0.10 eV. In all this
respect, the transport behavior of La$_{0.9}$Ca$_{0.1}$VO$_3$ is very
similar to that of LaVO$_3$, including very similar gaps in both
cases.  Quantitatively, however, the VRH appears to dominate over a more
extensive temperature range in the case of the Ca doped sample compared
to LaVO$_3$, as can be seen comparing the plots in the insets of Figs.
1 and 2.  This suggests that hole doped states arising from Ca
substitution in the place of La leads to hole doped states
in a similar energy range as that introduced by oxygen
nonstoichiometry in the parent compound. Larger extent of hole
doping in the former however leads to higher DOS at $E_F$ leading to
a manifestation of VRH more extensively, though the DOS at $E_F$ in
both cases are localized due to disorder effects.

Doping with $x \geq 0.2$ in La$_{1-x}$Ca$_x$VO$_3$ drives the system
into a metallic state. This is illustrated in Fig. 3 where we plot
the resistivity as a function of temperature for samples with $x$ =
0.2, 0.3, 0.4 and 0.5.  The plots clearly indicate metallic behaviour
of these compositions, except for an upturn of resistivity with
decreasing temperature below 100 K for $x$ = 0.2 and 0.3 samples
exhibiting a negative temperature coefficient of resistivity (TCR).
Similar behaviour has also been reported for La$_{1-x}$Sr$_x$VO$_3$
\cite{tokura}.  In spite of this increasing resistivity with decreasing
temperature below 100 K, the samples appear to have a 
finite $\rho (T \rightarrow 0$ K) .
This is better illustrated in
the inset where we plot the conductivities of these two samples as a
function of $T^{1/2}$ in the low temperature (negative TCR) region.
These two plots suggest that the low temperature conductivity $\sigma
(T)$ is well described by the relationship $\sigma (T) =
\sigma (0) + \beta T^{1/2}$, where $\sigma (0)$ is the finite
conductivity at $T = 0$ and the second term describes a square-root
behaviour. Such a temperature dependence of conductance is well known
in disordered metals in presence of strong electron-electron
interaction effects \cite{dismet1,dismet2}.  It is indeed reasonable
to expect that disorder will have some influence on the properties of
such samples. After all, such substituted samples are intrinsically
disordered due to the random substitution in the lattice. Moreover,
various disorder effects must be the driving force for the insulating
state obtained with finite but small level of substitution ($x <
0.2$), since there cannot be a Mott-Hubbard insulator or a band
insulator at arbitrary filling (or $x$). This view point is further
substantiated by the variable range hopping exhibited by the
insulating compositions. Thus, it is not surprising that the disorder
effects manifest itself even in the metallic phase near the critical
concentration in the form of a $ T ^{1/2}$ dependence of the conductivity
 at the low temperatures. Only deep into the metallic phase ($x \geq 0.4$) does
it appear that the random potentials are sufficiently screened out by
the metallic charge carriers and thus, there is no upturn of
resistivity at the low temperatures for such compositions (see Fig.
3).

While normal metallic systems are expected to show a $T^2$ dependence
of resistivity, recently it has been shown \cite{tokura} that the
resistivity of the closely related system La$_{1-x}$Sr$_x$VO$_3$ is
well described by a $T^{1.5}$ dependence.  This was attributed to
strong spin fluctuation effects \cite{spinflc1,spinflc2} with a
Curie-Weiss-like susceptibility near the antiferromagnetic wave
vector for these nearly antiferromagnetic metallic samples. We have
explored whether a similar behaviour is also found in the present
case. Thus, we plot the resistivity data as functions of $T^{1.5}$
and $T^2$ in Figs. 4a and 4b respectively for these metallic
compositions leaving out the part of the resistivity data with
negative TCR for $x$ = 0.2 and 0.3 at the lowest temperatures. These
results clearly reveal a systematic behaviour. 
For smaller values of $x$ (= 0.2 and 0.3), the resistivities at
higher temperatures ($ T > 100 $ K ) are better described by the
$T^2$ dependence and a $T^{1.5}$ dependence is clearly not consistent
with the data. With increasing $x$ (=0.4 and 0.5), however, $T^{1.5}$
dependence provides the best description for the experimental results
with clear deviations from a T$^2$ behaviour. This appears to be a
reversed trend compared to the case of La$_{1-x}$Sr$_{x}$VO$_3$
\cite{tokura}, where the $T^{1.5}$ behaviour was most dominant for
small values of $x$ close to the critical composition. While it is
not clear at present if the physics of the Sr and Ca substituted
LaVO$_3$ should be fundamentally different arising from the
structural differences referred to in the introduction, it is hoped
that single crystal data on both systems will resolve this issue in
the future.

The temperature dependence of magnetic susceptibility of the
insulating compositions, LaVO$_3$ and La$_{0.9}$Ca$_{0.1}$VO$_3$ have
been plotted in Fig. 5. In
both cases, 
there is a maximum in the susceptibility, typical of an
antiferromgnetic system; from these results, we obtain the N\'{e}el
temperature for 
LaVO$_3$ to be around 125 K and that of La$_{0.9}$Ca$_{0.1}$VO$_3$ to
be around 115 K. The further increase in susceptibility below $T_N$
in the case of La$_{0.9}$Ca$_{0.1}$VO$_3$ may be due to the presence of
localized isolated magnetic moments near the doped sites.
The magnetic susceptibilities of the metallic compositions are shown
in Fig. 6 for $x$ = 0.2, 0.3, 0.4 and 0.5. In every case the
susceptibility is nearly independent of temperature at the higher
temperature region with a distinct increase at the low temperatures.
Such an increase in the magnetic susceptibility at low temperature is
generally associated with the presence of localized magnetic
impurities, while temperature independent component arises from
Pauli-paramagnetism of the conduction electrons. Thus, we have
analyzed the data in Fig. 6 in terms of these two contributions.
These analysis yield Pauli paramagnetic susceptibilities
($\chi_{PP}$) of 958, 836, 828 and 728 $\times 10^{-3}$ emu/mole for
$x$ = 0.2, 0.3, 0.4 and 0.5 respectively.  Thus, there is a monotonic
decrease in susceptibility with $x$.  Moreover, in every case,
$\chi_{PP}$ appears to be enhanced by approximately a factor of two
compared to the free electron value, suggesting the presence of
correlation effects.

On the basis of the present results we summarize the evolution of the
electronic structure in La$_{1-x}$Ca$_x$VO$_3$ with composition as
depicted schematically in Fig. 7. Fig. 7a depicts the schematic
single particle excitation spectra for both the occupied (lower
Hubbard band, LHB) and unoccupied (upper Hubbard band, UHB) parts.,
separated by about 1.1 eV as suggested by the optical measurements
\cite{tokura}. Presence of oxygen non-stoichiometry in normally
prepared composition, LaVO$_{3+\delta}$, introduces hole doped states
primarily about 0.1 eV above the top of the LHB as shown in Fig 7b;
this is evidenced by the activated transport with a
characteristic gap of about 0.1 eV as shown in Fig 1. The low
intensity tails of these states overlap the top of the LHB and thus,
the Fermi energy, $E_F$, lies within a continuum of states. However, the
low intensity continuum states in the vicinity of the $E_F$ in Fig. 7b 
are localized due to disorder effects and
thus lead to the VRH dominated $T^{1/4}$ dependence of conductivity at
the lowest temperatures (see Fig. 1). Doping of Ca in place of
La in La$_{1-x}$Ca$_x$VO$_3$ with $x$ $<$ 0.2 (see Fig 7c) appears to
dope hole states further in a similar energy range as in the case of
LaVO$_{3+\delta}$ (Fig. 7b), thus, exhibiting similar temperature
dependent resistivities in both the samples (compare Figs. 1 and 2).
However, there are more continuum states at $E_F$ in presence of 10\%
Ca doping ($x$ = 0.1) compared to LaVO$_{3+\delta}$; thus, VRH in
the Ca-doped case dominates over a wider temperature range. With further
doping of Ca, the doped states become broader and more intense leading
to a significant overlap with the LHB of the undoped compound and 
consequently lead to the metallic ground 
state for $x$ $\geq$ 0.2 (Fig. 7d). Within
the metallic regime, the resistivity exhibits a $T^2$ behavior for
the lower values (0.2 and 0.3) of $x$, changing over to a $T^{1.5}$
dependence for the higher values (0.4 and 0.5), with the Pauli
paramagnetic susceptibilities exhibiting an enhancement of about two
compared to the free-electron values.

\section{Acknowledgements}

The authors thank G. Kotliar for useful discussions.
KM acknowledges the Council of Scientific and Industrial Research,
Government of India, for financial assistance.

\section{Figure Captions}

 Fig. 1 The resistivity of LaVO$_3$ as a function of the 
temperature. The open circles are the experimental data.  The solid
line shows the best fit assuming contributions from both  
variable range hopping and activated transport.
The inset shows the $ln~ \rho ~vs~ 1000/T$ and 
$ln~ \rho ~vs~ 1/T^{1/4}$ plots.

 Fig. 2 The resistivity of La$_{0.9}$Ca$_{0.1}$VO$_3$ as a function
of the 
temperature.  The open circles are the experimental data. The solid
line is the best fit assuming contributions from both 
variable range hopping and activated transport.
The inset shows the
$ln~ \rho ~vs~ 1000/T$ and $ln~ \rho ~vs~ 1/T^{1/4}$ plots. 

 Fig. 3 The resistivity of La$_{1-x}$Ca$_x$VO$_3$ for $x$ = 0.2, 0.3,
0.4 and 0.5 as a function of the temperature. The inset shows the $ln~
\sigma ~vs~ 1/T^{1/2}$ plots for $x$ = 0.2 and 0.3 in the low
temperature region ($T < 100$ K). 

Fig. 4 The resistivities of La$_{1-x}$Ca$_{x}$VO$_3$ for $x$ = 0.2,
0.3, 0.4 and 0.5 as functions of (a) $T^2$ in the upper panel and
(b) $T^{1.5}$ in the lower panel.

Fig. 5 The magnetic susceptibility of LaVO$_3$ (solid circles) and
La$_{0.9}$Ca$_{0.1}$VO$_3$ (open circles).

Fig. 6 The magnetic susceptibility of La$_{1-x}$Ca$_x$VO$_3$ for $x$ =
0.2, 0.3, 0.4 and 0.5.

Fig. 7 The schematic energy diagrams for
La$_{1-x}$Ca$_{x}$VO$_3$. (a) The case for $x$ = 0.0, (b) LaVO$_{3+\delta}$,
(c) $x$ $<$ 0.2, and (d) $x$ $\geq$ 0.2.

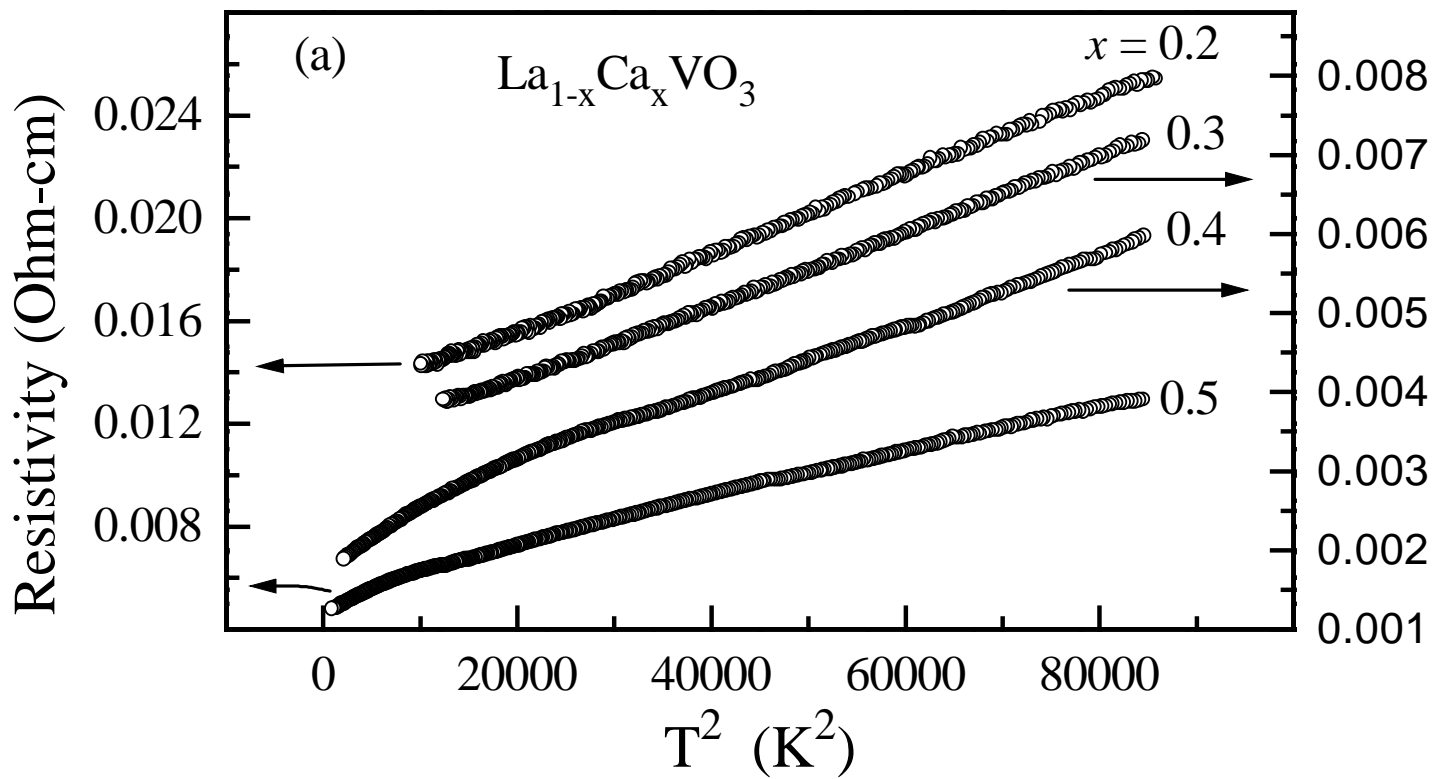
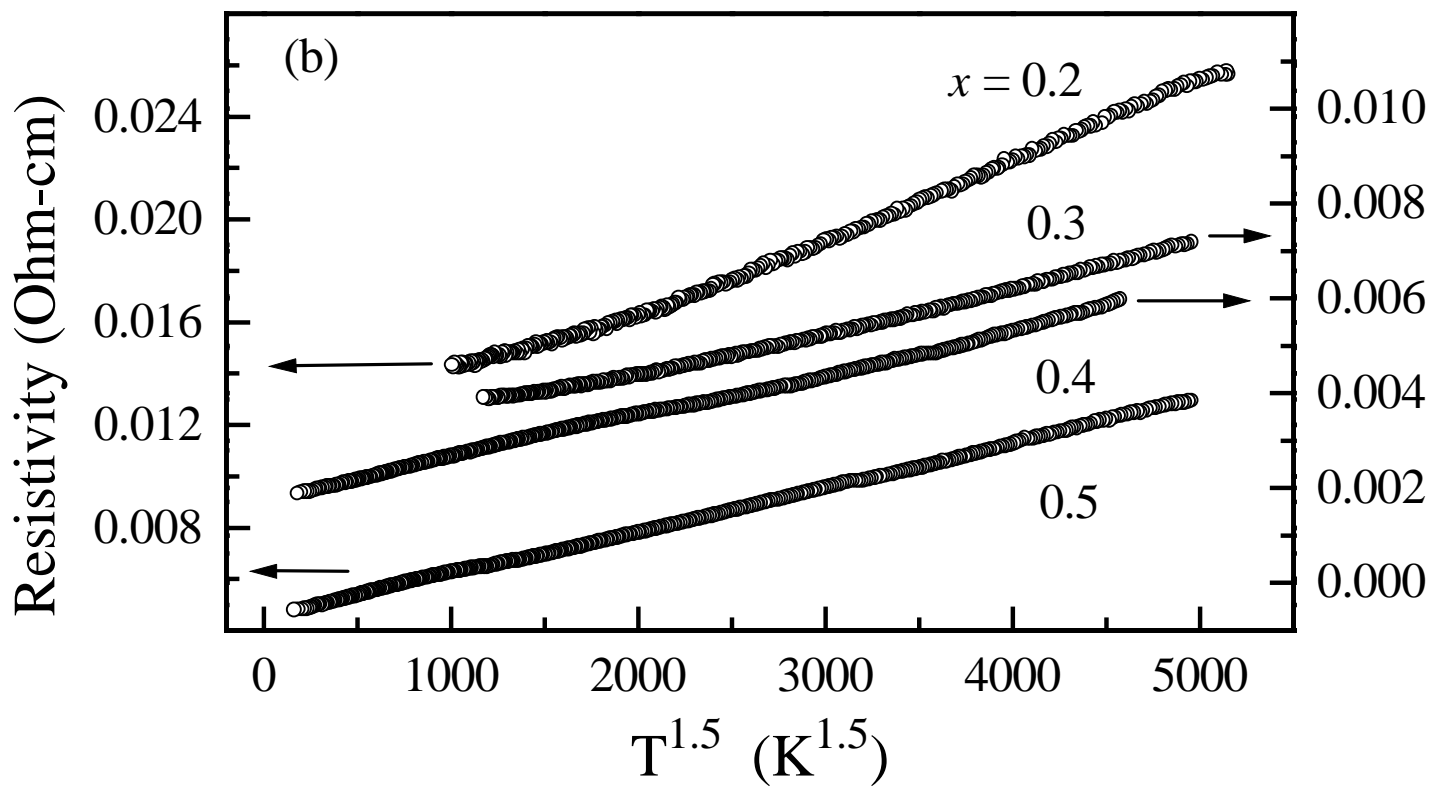
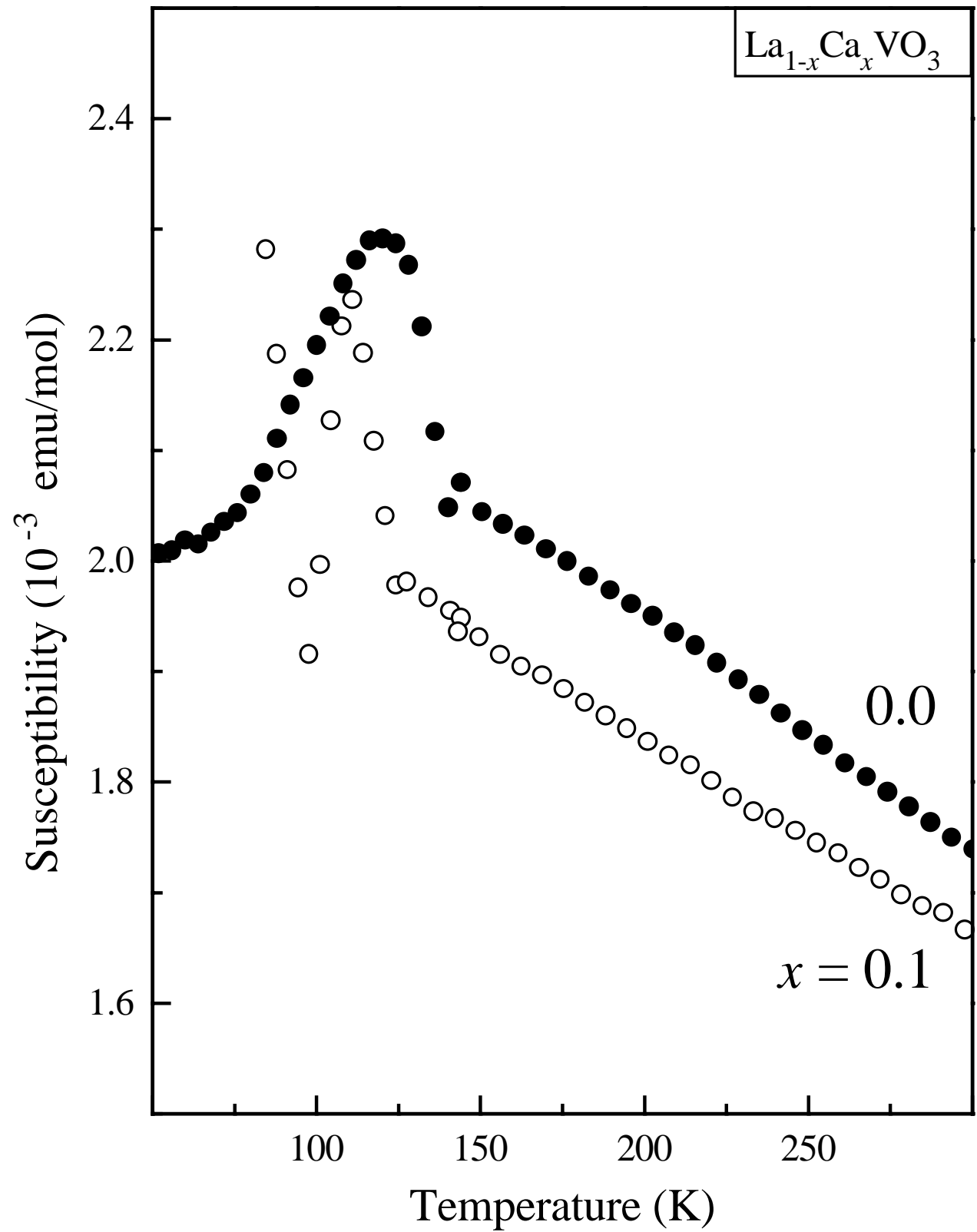
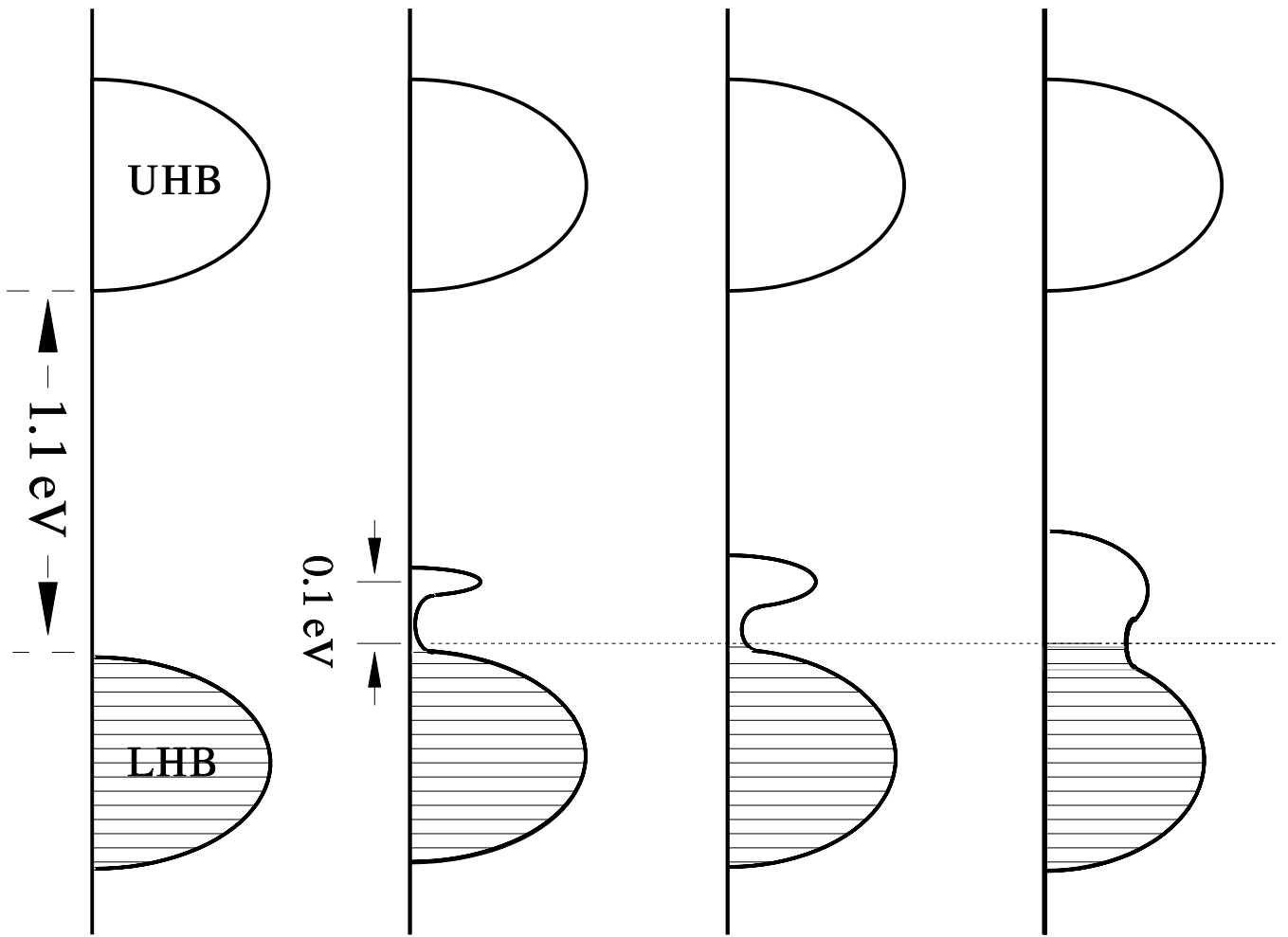
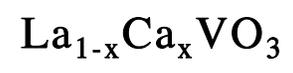
\end{document}